\documentclass[12pt]{article}
\usepackage{a4wide}
\usepackage{amssymb}
\usepackage{graphicx}
\begin{document}
{\renewcommand{\thefootnote}{\fnsymbol{footnote}}
\hfill  AEI--2005--133\\ 
\medskip
\hfill gr--qc/0508118\\
\medskip
\begin{center}
  {\LARGE Degenerate Configurations, Singularities and\\[2mm] the
    Non-Abelian Nature of
    Loop Quantum Gravity}\\
  \vspace{1.5em}
  Martin Bojowald\footnote{e-mail address: {\tt
bojowald@gravity.psu.edu},\\ new address: Institute for Gravitational
Physics and Geometry, The Pennsylvania State University, University
Park, PA 16802, USA}\\
  \vspace{0.5em}
  Max-Planck-Institut f\"ur Gravitationsphysik, Albert-Einstein-Institut,\\
  Am M\"uhlenberg 1, D-14476 Potsdam, Germany \vspace{1.5em}
\end{center}
}

\setcounter{footnote}{0}

\newcommand{\lP}{\ell_{\mathrm P}}
\newcommand{\tr}{\mathop{\mathrm{tr}}}
\newcommand{\sgn}{\mathop{\mathrm{sgn}}}
\newcommand*{\R}{{\mathbb R}}

\begin{abstract}
  Degenerate geometrical configurations in quantum gravity are
  important to understand if the fate of classical singularities is to
  be revealed. However, not all degenerate configurations arise on an
  equal footing, and one must take into account dynamical aspects when
  interpreting results: While there are many degenerate spatial
  metrics, not all of them are approached along the dynamical
  evolution of general relativity or a candidate theory for quantum
  gravity. For loop quantum gravity, relevant properties and steps in
  an analysis are summarized and evaluated critically with the
  currently available information, also elucidating the role of
  degrees of freedom captured in the sector provided by loop quantum
  cosmology. This allows an outlook on how singularity removal might
  be analyzed in a general setting and also in the full theory. The
  general mechanism of loop quantum cosmology will be shown to be
  insensitive to recently observed unbounded behavior of inverse
  volume in the full theory. Moreover, significant features of this
  unboundedness are not a consequence of inhomogeneities but of
  non-Abelian effects which can also be included in homogeneous
  models.
\end{abstract}

\section{Introduction}

By now, several discussions of the singularity issue have appeared in
the context of loop quantum cosmology \cite{LoopCosRev,WS:MB} and loop
inspired approaches \cite{BohrADM,Modesto} to quantum cosmology. As an
initial step, one often considers operators for inverse powers of the
determinant of the metric such as $(\det q)^{-1/2}$ as they are needed
for matter Hamiltonians and can be used for a first guess as to the
fate of classical singularities in a quantization.  In fact, many
models lead to bounded operators for inverse volume
\cite{Sing,InvScale,Bohr,BohrADM,Modesto,HWBlackHole} despite of the
classical divergence, following methods of full loop quantum gravity
\cite{QSDV}.  This is the case generically, for instance, in isotropic
cosmology \cite{IsoCosmo,Bohr} where bounded inverse volume
expressions play a central role in the non-singular evolution as it
follows from a difference equation for the wave function.

The issue is already different, however, in anisotropic models
\cite{HomCosmo,Spin} where the same quantization gives matter
densities or curvatures which are not necessarily bounded when
considered on full minisuperspace. In those cases, a non-singular
evolution exists nonetheless, which demonstrates that, by itself,
boundedness of kinematical operators such as inverse volume is not
essential and that its role can only be seen when combined with
dynamical information such as that given by the Hamiltonian
constraint. There are also other cautionary points one has to keep in
mind when using inverse volume properties to understand the
singularities.  For instance, implicit in using inverse volume for
statements about singularities is the classical relation between
curvature divergence and inverse metric components. Even if all
inverse powers of the volume could be shown to be finite at the
quantum level, one would still have to discuss their relation to
space-time curvature, so again dynamical properties, which is more
complicated.  While classically, at least in isotropic cases,
extrinsic curvature is proportional to an inverse power of the volume,
this follows only after equations of motion are solved. Thus, also at
the quantum level one has to use equations of motion to relate
properties of inverse volume expressions to the singularity issue.

Nevertheless, in order to obtain a more complete picture it is
important to analyze the behavior of inverse volume in different
models and also, as far as possible, in the full theory. Recently
\cite{BoundFull}, partial information from the full theory has been
derived for gauge-invariant 3-vertices of spin network states which
are always annihilated by the volume operator \cite{AreaVol,Vol2}.
Here, inverse volume operators have, as functions of the edge spins,
unbounded kinematical expectation values in 3-valent vertex
states. The new situation, unlike any example used in isotropic,
anisotropic or even some inhomogeneous models so far
\cite{IsoCosmo,HomCosmo,Spin,SphSymmVol,SphSymmHam}, is that
unboundedness occurs already on states which have zero volume
eigenvalue. (Which still is an improvement compared to the classical
behavior which would predict diverging inverse volume. That this is
not happening in the loop quantization is a direct consequence of the
techniques introduced in \cite{QSDI,QSDV}.)  While it would be
possible to construct operators with similar properties in models, in
loop quantum cosmology they would break parity invariance of the
theory.  (Parity, i.e.\ ${\rm sgn}\det E$ for the densitized triad
$E^a_i$, has not been studied in the full calculations done so far
since only configurations of vanishing determinant are considered.)

When a difference between symmetric models and the full theory is
discovered, the first reflex is often to blame it, without further
corroboration, on a breakdown of the minisuperspace approximation. So
also in this case, e.g.\ inhomogeneities have quickly been named as a
potential culprit.  However, there are many technical and conceptual
differences between models and the full theory, and so several reasons
for discrepancies can be imagined. For a reasoned and informed
judgment a deeper investigation is thus warranted. To illustrate this
we mention atomic spectra as an example, where characteristic
properties of the hydrogen atom obtained from a spherically symmetric
potential remain intact but only become more complicated in details if
realistic interactions are switched on. In particular, the atom
remains stable unlike its classical counterpart. This is, of course,
only true if interactions are restricted (in this case from other
theoretical investigations or observations) and not allowed to be
arbitrary. Under arbitrary interactions, properties can certainly
change dramatically, and indeed do so for atoms of high central charge
where the electromagnetic interaction is stronger.  Similarly, in
quantum gravity it is not sufficient to look at arbitrary degenerate
geometrical configurations unless their meaning is known.  There must
be an analog of switching on the {\em correct} gravitational
interaction, e.g.\ between inhomogeneities, which means that
constraint equations or observables need to be considered.

While in \cite{BoundFull,BoundCoh} the main aim was to derive and
point out potential differences (also emphasizing, as oftentimes
before in loop quantum cosmology, that dynamical information has to be
included), here we analyze and contrast different geometrical
configurations so as to facilitate developing a better idea for the
relation between models and the full theory. We will then first review
dynamical information about the singularity issue and discuss
properties of inverse volume. New considerations start with
Sec.~\ref{Abel}, which contains aspects of the regularization
procedures of loop quantum gravity and differences to analogous
procedures in an Abelian truncation which play a role for geometrical
operators but have not been recognized thus far. This indicates new
sources of contributions to effective equations from non-Abelian
properties which can play a role for cosmological phenomenology. An
analysis of inverse volume operators in different models and
truncations to be discussed in Sec.~\ref{Deg} reveals a possible
technical origin of bounded behavior which, however, is not directly
related to physical information such as degrees of freedom. After
comparing the behavior in models with the full theory and its Abelian
truncation in Sec.~\ref{Comp}, we discuss in Sec.~\ref{Evol} aspects
of dynamics which in any situation ultimately has to be studied for a
conclusion about singularity removal. We will find potential dynamical
effects of those properties of the full theory studied in
\cite{BoundFull}.

\section{General properties of singularity removal}

The main problem posed by a singularity is the breakdown of dynamical
equations at such a point.  In any complete discussion of quantum
removal of classical singularities one thus has to address three
points: 
\begin{enumerate}
\item First, one has to isolate potential singularities, i.e.\ find
conditions under which singularities must occur for the classical
equations. 
\item One then has to determine the quantum evolution
equations for the dynamics around those classically singular states.
\item Finally, this dynamics has to be shown not to break down when
quantum behavior is taken into account.
\end{enumerate}
All three steps, even the first purely classical one, are too
complicated to approach in full generality but can be dealt with in
symmetric models which classically also show a wealth of different
singularities. When comparing with the full theory, one then focuses
so far on the issue of boundedness of quantities related to curvature,
motivated by the fact that often curvature diverges at classical
singularities. More importantly, such expressions, e.g.\ the inverse
square root of the determinant of the metric, also enter matter
Hamiltonians and thus appear in parts of the dynamical
equations. Boundedness then is still not required but may simplify the
analysis. Moreover, only the second and third steps are avoided while
the first one of isolating singularities is still present at full
force. In this section, we will collect the currently available
information about evolution and inverse volume before entering more
detailed aspects of a comparison between symmetric and the full
situations later on.

\subsection{Evolution}
\label{General}

The boundedness of inverse volume operators on superspace has proven
not to be relevant for the general mechanism of singularity removal in
models of loop quantum gravity as long as there is a well-defined
expression at all.  The isotropic case, where these developments have
started, is however quite special which has led to some confusion and,
occasionally, an overemphasis of certain aspects. There is only one
parameter characterizing an isotropic spatial geometry and thus only
one way to approach the classical singularity on minisuperspace. In
this case, a well-defined behavior of inverse powers of volume would
imply boundedness which indeed is realized automatically
\cite{InvScale}.  The most crucial aspect of non-singular behavior,
however, is a unique extension of evolution beyond the classical
singularity which is {\em not} guaranteed even for bounded inverse
volume (see, e.g., the discussion in \cite{IsoCosmo}). On the other
hand, this extension of evolution has been generalized to
non-isotropic cases (making use of difference equations representing
the Hamiltonian constraint in a triad representation) even when
inverse powers of volume or curvature components are not bounded on
the respective superspaces.

The key reason for using models is that for them the singularity
structure is often clear and the first task above can indeed be
performed. Other simplifications may then arise which facilitate a
direct analysis of the evolution but are not necessarily crucial for
the result.  For the evolution one has to specify initial values and
usually also boundary values for the wave function in the {\em
non-singular} part of superspace, such that one obtains a well-posed
initial value problem. In non-singular situations, these data uniquely
give the solution to the constraint not only on one classically
connected part of superspace, but also on other parts separated by
classical singularities. In this sense, one can evolve beyond the
classical singularity and the singular boundary is removed.

Note that, although the well-posedness of initial/boundary value
problems for difference equations does play an important role, this is
not simply an issue of counting the number of solutions through
initial values and that the existence of an extension is
non-trivial. Singular evolution in this setting arises when values of
the wave function at the singular boundary are not determined (due to
vanishing leading coefficients in a difference equation) but would be
needed for evolving further. One can, of course, simply include those
values as data to be specified, but this does not solve the
singularity problem. It would even be possible classically, e.g.\ in
isotropic cosmology where one can specify the scale factor and matter
fields at, say, $t=1$ as well as $t=-1$ when the singularity is at
coordinate time\footnote{We use here an argument in coordinate time
since it is usually used in the classical situation. One could equally
use internal time.}  $t=0$, and then solve the Friedmann equation with
both sets of initial values and simply glue the solutions together. In
the classical as well as quantum case this would give a (non-unique)
evolution beyond the classical singularity but would only remove a
breakdown of evolution by putting in missing information by hand. For
singularity removal, it is necessary to demonstrate a complete
extension of wave functions given values in only one connected
non-singular part of superspace.

Crucial ingredients are a quantum equation for the wave function on a
suitable configuration space which requires an explicit construction
from quantum gravity and involves analytical techniques. This equation
is often called ``evolution equation'' even though it is not necessary
to have a global time evolution picture. One then needs to identify
classically singular parts of the configuration space and suitable
transversal directions along which classical evolution breaks
down. Transversality means that the singularity is not simply
approached asymptotically but is reached after a finite coordinate
distance on configuration space.  These directions would, in an
evolution picture, be parameterized by internal time variables and
thus a singularity will be reached after a finite amount of internal
time has passed. It is then possible to ask and investigate what can
happen in a more complete and non-singular theory after that time, or
beyond the classical singularity.

In this step one uses classical gravity and geometry. All ingredients
then have to be combined in order to see if a wave function can be
extended uniquely along the transversal directions, which would
correspond to evolution through the classical singularity (see also
\cite{SphSymmSing}). One can conclude non-singular behavior only if
the extension is possible for any state or, if one has information on
the physical inner product, any state relevant for the physical
Hilbert space. In this way, the explicit construction of observables
and a reconstruction of the corresponding geometry is by-passed, but
nonetheless models can be concluded to be {\em physically}
non-singular: if an extension for the wave function to a new region is
known, one can extract information from it on both sides of a
classical singularity. This is in particular true for loop quantum
cosmology where kinematically both sides differ only by their
orientation. If extracting an effective geometry at one side is
well-defined, as is required for the correct semiclassical limit, one
can do so equally at the other side for any parity invariant
observable.

\begin{figure}
\begin{center}
\includegraphics[width=10cm]{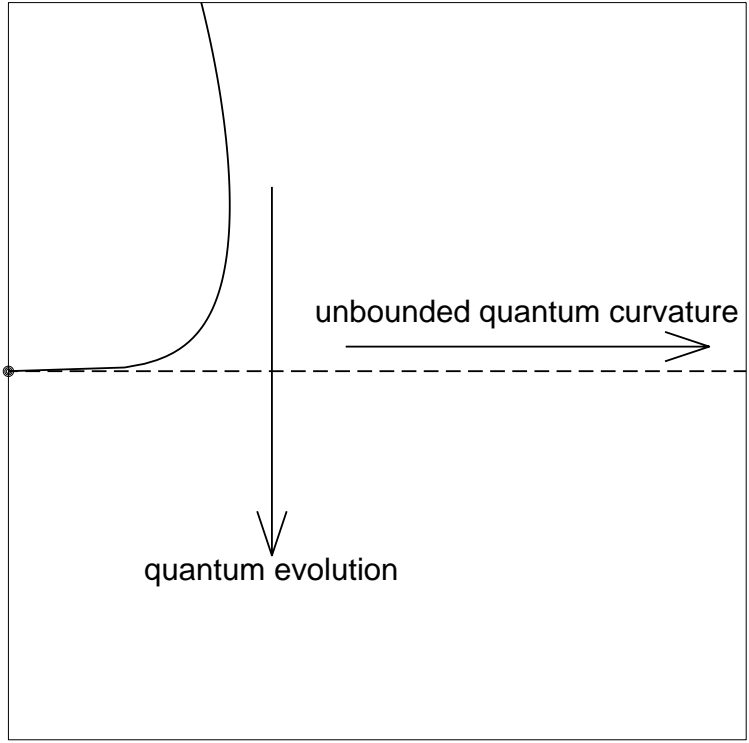}
\caption{Illustration of the quantum behavior for wave functions on
  the 2-dimensional minisuperspace of an anisotropic model compared to
  the classical situation (see also \cite{HomCosmo,BHInt}). This is
  only a sketch of part of minisuperspace extracted from models
  studied so far. Any classical trajectory on minisuperspace (solid
  curve as one example) runs into the classical singularity (circle).
  Classical expressions for curvature diverge at the dashed line which
  contains the classical singularity.  Quantum expressions for
  curvature, however, are well-defined on the dashed line and may, if
  unbounded, grow in a parallel direction (corresponding to increasing
  spin labels of degenerate states, or some triad components becoming
  large). The quantum evolution proceeds in a direction transversal to
  the classically singular dashed line and uniquely extends a wave
  function to the lower part of minisuperspace (obtained through
  orientation reversal) not reached by the classical evolution. This
  demonstrates the importance of knowing the singularity structure on
  superspace, and the irrelevance of boundedness on or close to
  degenerate states. Unboundedness can only matter if one approaches
  the classical singularity asymptotically, for growing minisuperspace
  coordinates to the right in this figure. Such an asymptotic
  trajectory parallel to the classically singular line would simply
  amount to an internal time variable chosen in an inappropriate
  manner and thus reflects only the analog of a coordinate
  singularity.}
\label{fig}
\end{center}
\end{figure}

What is not strictly necessary but often useful for an explicit
analysis are the existence of a triad representation, a global time
variable, or bounded inverse volume on superspace (in general, it is
enough that inverse volume remains well-defined along any transversal
direction relevant for the quantum evolution; see Fig.~\ref{fig}). If
a triad representation exists, one obtains a more intuitive picture
through evolution by a difference equation.  

Bounded inverse volume or curvature can be helpful since any state
will be in their domain of definition. The action of the matter
Hamiltonian can then be computed in each iteration step of a
recurrence scheme. If such an operator is not bounded and only densely
defined, it could happen that one needs to act on a state arising
during the evolution on which the operator is not defined.  This could
be a reason for a breakdown of evolution related to unbounded
operators. With such a property the analysis would be more
complicated, but it does not immediately imply singular behavior. The
dynamical behavior would have to be analyzed in a detailed manner, and
all initial values leading to such a breakdown must be identified.
Even if one would obtain a unique extension for generic initial values
(which is likely because operators in any case have dense domains),
it would be impossible to claim singularity freedom unless one can
show that the singular states are removed from the physical Hilbert
space. If, on the other hand, arbitrary initial values lead to a
uniquely extended state, it does not matter for the singularity issue
how exactly the physical Hilbert space is obtained from the space of
solutions to the constraint.

\subsection{Inverse volume spectra: Models and the full theory}
\label{InvVol}

Having discussed the dynamical aspects, we briefly recall now which
operators have been studied for quantizations of $(\det q)^{-1/2}$ and
their main properties. All these constructions follow the general
scheme of classically expressing inverse metric components through
Poisson brackets between connection components $A_a^i$ and a positive
power of volume based on
\cite{QSDI,QSDV}
\begin{equation} \label{ident}
 2\pi\gamma G
 \epsilon^{ijk}\epsilon_{abc} \frac{E^b_jE^c_k}{\sqrt{\left|\det
   E\right|}} \sgn\det E= \{A_a^i,\int\sqrt{\left|\det E\right|}
  \mathrm{d}^3x\}\,,
\end{equation}
($\gamma$ is the Barbero--Immirzi parameter \cite{AshVarReell,Immirzi}
and $G$ the gravitational constant). Instead of the spatial metric
$q_{ab}$, a densitized triad $E^a_i$ is used which is canonically
conjugate to the connection \cite{AshVar,AshVarReell}.  This relation
is then quantized using holonomies of $A_a^i$, the volume operator,
and turning the Poisson bracket into a commutator. The resulting
operator is densely defined, and in some models it was found to be
even bounded. A first explicit expression for a spectrum was computed
in the isotropic context \cite{InvScale} where, e.g.,
\begin{equation} \label{Iso}
 (\left|\det E\right|^{-1/2})_{\mu} = \left(4\gamma^{-1}\ell_{\mathrm{P}}^{-2}
 (\sqrt{V_{\mu+1}}-\sqrt{V_{\mu-1}}\,)\right)^6
\end{equation}
on states $|\mu\rangle$ with a single label $\mu\in\R$ whose
geometrical meaning is given through the densitized triad component
$p$ with eigenvalue $p_{\mu}=\frac{1}{6}\gamma\lP^2\mu$
($\ell_{\mathrm{P}}=\sqrt{8\pi\hbar G}$ is the Planck length and
$V_{\mu}=|p_{\mu}|^{3/2}$ a volume eigenvalue).  The absolute value of
$\mu$ thus determines the volume, while its sign is spatial
orientation encoded geometrically in the handedness of the triad. The
fact that the isotropic inverse scale factor is bounded has played a
major role in further developments leading to a demonstration of
singularity-free evolution in various models.

Already in anisotropic but still homogeneous models the situation is
different since volume can become small even if some metric components
are large. The single quantum number $\mu$ is replaced in diagonal
homogeneous models \cite{HomCosmo} by three labels $\mu_I$,
$I=1,\ldots,3$, which determine the three diagonal components of a
densitized triad by
$p^I_{\mu_1,\mu_2,\mu_3}=\frac{1}{2}\gamma\lP^2\mu_I$, the volume
by $V_{\mu_1,\mu_2,\mu_3}=(\frac{1}{2}\gamma\lP^2)^{3/2}
\sqrt{|\mu_1\mu_2\mu_3|}$ and the orientation by
$\sgn(\mu_1\mu_2\mu_3)$. Volume can then become small not only when
all $\mu_I$ go to zero but also when, say, two of them go to zero
while the third diverges at a suitable rate. This can, and generically
does, lead to unbounded behavior of inverse volume eigenvalues at
smaller volume unless cancellations occur.

For an inverse volume as above, or more generally a power of the
volume parameterized by $r>0$, one arrives at an expression of the
form
\begin{equation} \label{Vdiag}
 (V^{6r-4})_{\mu_1,\mu_2,\mu_3}=({\textstyle\frac{1}{2}}\gamma\lP^2)^{9r-6} 
 F_r(\mu_1)|\mu_2\mu_3|^{r}
 F_r(\mu_2)|\mu_1\mu_3|^{r} F_r(\mu_3)|\mu_1\mu_2|^{r}
\end{equation}
where $F_r$ is a function behaving as $F_r(\mu)\sim \mu^{r-2}$ for
$\mu\gg1$ but cutting off the classical divergence at small $\mu$.
This is obtained using, analogously to Eq.~(2.5) of \cite{BoundFull}
and its generalization to $r\not=\frac{1}{2}$, the classical relation
\[
V^{6r-4}=|p^1p^2p^3|^{3r-2}= (4\pi\gamma Gr)^{-6}
\left(\prod_{I=1}^3\{c_I,V^r\}\right)^2
\]
where $c_I$ are homogeneous connection components conjugate to the
triad components $p^I$, $\{c_I,p^J\}=8\pi\gamma G\delta^J_I$.  One
quantizes the right hand side by expressing the connection components
through ``holonomies,''\footnote{On the relevant Hilbert space of
  functions on the Bohr compactification of the real line, the
  components $c_I$ are not represented directly but only their
  exponentials $e^{i\delta c_I}$ in a manner not continuous in
  $\delta$ \cite{Bohr}.}  e.g.\ $\{c_I,V^r\}=
2\tr(\tau_Ie^{c_{(I)}\tau_I} \{e^{-c_{(I)}\tau_I},V^r\})$ with SU(2)
generators $\tau_I=-\frac{i}{2}\sigma_I$, using the volume operator
and turning the Poisson bracket into a commutator.  This leads to
functions
\[
 F_r(\mu)=r^{-2}(|\mu+1|^{r/2}-|\mu-1|^{r/2})^2\,.
\]

Since there are still positive powers of $\mu_I$ in (\ref{Vdiag}),
unboundedness could be possible when, e.g., $\mu_1$ becomes large.
While in this case we can use the large-$\mu$ behavior of $F_r$ and
see that $\mu_1$ does not lead to unboundedness for an inverse power
(i.e.\ $r<\frac{2}{3}$), other quantizations or different curvature
components, where not a product as above occurs but instead a sum,
lead to unbounded behavior at small volume. With the above
quantization, however, inverse volume or curvature operators are
always {\em bounded, and in fact zero, on states of zero volume\/}
since $F_r(0)=0$.

The latter property is not realized in the full theory where a
non-zero and unbounded expectation value in 3-valent vertex states,
equation (4.19) of \cite{BoundFull}, has recently been derived. The
expression is again available for any positive value of $r$ if one
only replaces the powers $\frac{1}{4}$ of that formula by
$\frac{r}{2}$.  It can thus not only be used for inverse volume
($r=\frac{1}{2}$) but also for other powers such as the volume itself
($r=\frac{5}{6}$). In all these cases the function of the three spins
$j_I$ is {\em unbounded on zero-volume eigenstates}, showing clearly a
difference to homogeneous models. This behavior on zero-volume states
has been used in \cite{BoundCoh} to argue that the singularity
structure of full loop quantum gravity may not be captured reliably by
models.

What should make one suspicious, though, is the fact, unnoticed in
\cite{BoundFull}, that not only the inverse volume ($r=\frac{1}{2}$)
is unbounded on zero-volume states but also the determinant of the
metric ($r=\frac{5}{6}$), which can be taken as the volume of the
vertex. Thus, even though the volume eigenvalue is zero on any such
state, there is a quantization of the volume which can be arbitrarily
large.\footnote{One can change the quantization by introducing a sign
  factor for the sign in (\ref{ident}), which would give zero
  expectation values of the new operator in any zero-volume state
  \cite{TT}. However, this property would equally apply to inverse
  powers which then also become zero and thus bounded. Though maybe
  valuable, we will not consider this quantization choice further in
  this paper because it would remove the differences between models
  and the full theory related to boundedness of inverse volume on
  zero-volume states.} This should not come as a big surprise since
any such state must be considered deeply quantum and quantization
ambiguities can have large effects.  But it also shows that the
geometrical, let alone dynamical, meaning of these configurations must
be understood better before the results can be used for definitive
conclusions. In particular the geometrical meaning of the spin labels
around a vertex should be better known, similarly to the situation in
homogeneous models where the parameters can directly be associated
with quantized metric components. Unlike in models, a suggestive
identification of those states of zero volume, or other states
containing those vertices, as ``big bang states'' is premature.

One particularly important aspect of a geometrical configuration used
for establishing non-singular evolution in models is parity. Since
loop quantum gravity is based on a triad formulation of gravity, the
orientation of space (not through the background manifold but through
the basic variables of the theory) is encoded in $\sgn\det E$. The
classical formulation is not much different from a metric formulation
since the two orientations are not connected by physical trajectories
but instead separated by singularities. This changes in loop quantum
cosmology where the quantum evolution equation connects a wave
function uniquely from one parity side to the other. For this, a
region of reversed parity is required since otherwise it would not be
possible to extend evolution. This is the main scheme for non-singular
evolution, and its main technical ingredient is the parity degree of
freedom and the structure of the evolution equation
\cite{Sing,HomCosmo,SphSymmSing}.
It is clear from those considerations that it is not enough to
consider only zero-volume states in isolation, i.e.\ static
properties. The main part to a conclusion of singularity removal is a
consideration of transitions where zero-volume states appear, which
can only come from whatever evolution equation one is using. These are
thus dynamical properties, which are crucial for any singularity
statement. After providing more details about degenerate
configurations, we will come back to dynamics in Sec.~\ref{Evol}.

\section{Approximations in the full theory}

Since exact calculations in the full theory are usually quite
involved, approximations other than symmetry reduction are necessary
if a wider picture is to be developed. In the present context, it can
be helpful to employ an Abelianization by replacing the structure
group SU(2) by ${\rm U}(1)^3$ and using suitable definitions of the
volume operator and other objects \cite{QFTonCSTII}. Many calculations
then simplify, which has already been used to approach zero-volume
states in a coherent state and to demonstrate that at least in this
case expectation values of inverse volume can remain bounded along
isotropic classical trajectories \cite{BoundCoh} (crucially relying on
classical and {\em symmetric} properties such as how an isotropic
singularity is approached). As we will show now, however, the
boundedness issue for inverse volume operators is quite different in
${\rm U}(1)^3$ compared to SU(2). Nevertheless, ${\rm U}(1)^3$
calculations are not only much simpler but also provide additional
information which is of interest for comparison with symmetric models.

\subsection{Abelianization vs.\ Homogeneity}
\label{Abel}

In fact, many symmetric models turn out to be related to ${\rm
  U}(1)^3$-vertices.  The labels $\mu_I$ of a diagonal homogeneous
model, for instance, can be viewed as corresponding to a 6-valent
${\rm U}(1)^3$-vertex with edges given by straight lines entering and
leaving the vertex and having charges $(\mu_1,0,0)$, $(0,\mu_2,0)$
and $(0,0,\mu_3)$, respectively. Different holonomies in the model are
still non-commuting and thus take SU(2)-properties into account, but
as far as geometrical expressions such as the volume are concerned the
model can be formulated in an effectively Abelianized manner, which is
the main reason for simplifications. (Non-Abelian properties, however,
enter the dynamics as given by the Hamiltonian constraint.)  Looking
at ${\rm U}(1)^3$-configurations thus provides means to distinguish
between different aspects of a symmetry reduction and to see which of
them are responsible for special behavior in models.  However, since
introducing ${\rm U}(1)^3$ also presents an approximation to the full
theory, one must as well understand its range of validity and possible
deviations to full SU(2)-calculations.

There is indeed one prominent place in our context where the
non-Abelian nature plays a role. In the rewriting procedure one makes
use of relations of the form
\begin{equation} \label{commut}
 \epsilon\dot{e}^a\{A_a^i,V\}\approx 2\tr(\tau_i h_e\{h_e^{-1},V\})+
 O(\epsilon^2)
\end{equation}
where $h_e$ is a holonomy along an edge $e$ of parameter length
$\epsilon$ and $V$ the volume of a region ($\tau_i$ are generators of
the structure group). The left bracket can thus be expressed by
holonomies up to terms of order $\epsilon^2$ which vanish in a
continuum limit. Alternatively, at fixed $\epsilon$ the right hand
side can be seen as an approximation or regularization of the left
hand side. Irrespective of the interpretation of $\epsilon$-terms, the
quantization does not depend on this parameter. The Poisson bracket
gives then rise to a commutator in a vertex contribution to an
operator, and the holonomy can be elongated or shortened at the end
away from the vertex without changing the action on states.

As mentioned before in homogeneous models (see Sec.~\ref{InvVol}), the
application of identities such as (\ref{commut}) is necessary because
the Hilbert space obtained through a background independent
quantization does not allow the action of individual connection
components but only of holonomies. This is a characteristic property
of the quantization and is present irrespective of details of a
regularization or its interpretation. It is one possible origin of
correction terms in, e.g., an effective Hamiltonian compared to the
classical expression just as higher curvature terms often arise in
effective actions \cite{DiscCorr,SemiClassEmerge,Perturb,Josh}. For
the arguments to follow it is important to notice that there are two
different kinds of corrections on the right hand side of
(\ref{commut}), which usually are both subsumed in the symbol
$O(\epsilon^2)$: First, one has to approximate the connection
component in a single point by a suitable edge integral and, secondly,
that integral is exponentiated.\footnote{In non-Abelian cases these
are not two independent and subsequent steps in a computational
procedure; they are rather mixed up in the path ordered exponential
entering the holonomy.}  If the edge length is small, both
modifications imply only small changes of the indicated order.

What is different in these two types of corrections is their
dependence on the phase space variables. The left hand side of the
equation, even though it contains connection components in the Poisson
bracket, depends only on triad components after the bracket is
computed. Similarly, if the connection component is replaced by an
integral which is still linear in $A_a^i$, the bracket is just
evaluated in different points along the edge and still depends only on
triad components. When the integral is exponentiated, however, the
situation changes since in a non-Abelian theory the cancellation
between $h_e$ outside the bracket and $h_e^{-1}$ inside is not
complete: we have, schematically, $\delta h_e/\delta A_a^i(x)=h_0(x)
\tau_i h_1(x)$ where $h_{0/1}(x)$ denote the holonomies running from
the starting point of the edge to $x$ and from $x$ to the end point,
respectively. Only the contribution from $h_1$ cancels in
(\ref{commut}), but we are left with $h_0\tau_ih_0^{-1}$. Again, this
gives only small correction terms, but {\em they do depend on
connection components}, not just on the triad.

In an Abelianization, on the other hand, the second type of correction
terms does not appear because now the cancellation between $h_e$ and
$h_e^{-1}$ is complete. There are still $O(\epsilon^2)$-corrections of
the first kind, but they are independent of connection components.
Thus, in the Abelian case a commutator as vertex contribution is a
more direct quantization of the classical expression on the left hand
side of (\ref{commut}) which is not the case in a non-Abelian
framework where regularization has stronger effects.  A further
consequence is that the commutators obtained in a non-Abelian setting
do not commute with the volume operator even though they are supposed
to quantize objects which classically do not depend on connection
components.  Again, this can be seen as resulting from correction
terms as above, which do depend on the connection.  Indeed, in Abelian
quantizations such commutators commute with the volume operator, even
though holonomies appear in intermediate stages.

The two different correction terms also have different implications
for homogeneous models. The first one, originating in replacing
connection components by integrals, is clearly a consequence of a
field theory. It does not arise in homogeneous models where we can
simply take exponentials such as $e^{c_{(I)}\tau_I}$ as used before.
Thus, if the first correction would be crucial for the properties of
\cite{BoundFull}, it would clearly be an indication that homogeneous
models do not reliably capture the full situation. The second type of
correction term, on the other hand, is a consequence of non-Abelian
behavior. It disappears in an Abelianization irrespective of whether
it is introduced as a truncation to the full theory or effectively a
consequence of a particular symmetry reduction. Moreover, it even
occurs in non-diagonal homogeneous models where not all derivatives
with respect to connection components of, e.g., holonomies of the form
$e^{c_{(I)} \Lambda_I^i\tau_i}$ (as explained in more detail in
Sec.~\ref{Models}) are proportional to the same internal direction
$\Lambda_I^i\tau_i$. These properties, which we will discuss further
in what follows, thus allow one to {\em distinguish between non-Abelian
effects and effects of inhomogeneities}.

The above remarks on regularization are by no means saying that the
quantization would be wrong or suspect.  The classical limit is not
affected since the relation is valid in regions of phase space where
semiclassical behavior is expected. In this limit, one obtains similar
behavior of Abelian and non-Abelian calculations (which in some cases
can also be demonstrated explicitly \cite{CohState}). But when regimes
close to potential singularities are studied, one should pay utmost
attention to such correction terms.  In fact, the rewriting procedure
introduces such corrections which are expected to be strong close to
singularities and can significantly influence even qualitative
behavior. In addition, quantization ambiguities appear since rewriting
is possible in different ways.\footnote{Quantization ambiguities have
  been studied in detail in diagonal homogenous models and found not
  to be essential for qualitative properties \cite{Ambig,ICGC}. The
  new ambiguities arising in a non-Abelian context, however, are of a
  different type and still have to be analyzed.}

One example for such quantization ambiguities in action is the already
observed discrepancy, based on \cite{BoundFull}, between unbounded
behavior of a full quantization of $\sqrt{\det q}$ in a vertex even
though the basic volume operator annihilates that state. Since both
operators correspond to the same classical expression, this different
behavior is caused by a quantization ambiguity which can be traced
back to the relation (\ref{commut}) discussed above. In this case, it
is clear that the volume operator should be considered more basic
since it can be quantized directly using only fluxes
\cite{AreaVol,Vol2}, and as such it is indeed essentially unique
\cite{Flux}.  Then again, the fact that it annihilates all 3-vertices
is very special, resting solely on gauge invariance, and so
considering only 3-vertices is potentially misleading. Such properties
of different operators quantizing volume show that a geometrical
interpretation of vertex configurations can be complicated. Unless
other information is available, it is safest to use only regimes where
the volume operator is close to other quantizations of $\sqrt{\det
  q}$, which in the case of 3-valent vertices means small spins. As
for inverse volume operators, they are clearly bounded on any bounded
range of spins even though the volume eigenvalue vanishes. Moreover,
on those states, the quantized $(\det q)^{-1/2}$ is smaller than
$\sqrt{\det q}$ even though this corresponds to states of small
volume. There is thus a cut-off of classical divergences on degenerate
vertices of the full theory.

In light of the two types of correction terms, the difference of the
two volume operators on 3-vertices must be a consequence of the second
type of corrections since it does not occur, for the same valence of a
vertex,\footnote{On vertices of higher valence, as dicsussed later, it
  can happen also in the Abelian case that states annihilated by the
  basic volume operator are not annihilated by commutators. This is an
  effect of cancellations in the volume operator occurring at higher
  valence, which can annihilate states even for large edge labels.
  Also here, inhomogeneities are not relevant which can be seen from
  the fact that for individual commutators compared to flux operators
  no crucial ambiguities arise even on higher-valent Abelian
  vertices.} in an Abelian truncation (see the following subsection).
It is thus not a consequence of inhomogeneities. It is, rather, a
consequence of additional correction terms which only appear in a
non-Abelian setting and can lead to more distinguishing features
between otherwise similar operators. The same conclusions then apply
to the behavior noticed in \cite{BoundFull} where commutators such as
inverse volume or the alternative volume are unbounded on states which
are annihilated by the basic volume. It is then crucial to understand
which operator among different choices is relevant for the
identification of singular states.

Fortunately, the presence of quantization ambiguities does not
preclude definitive conclusions. In the present case, where different
quantizations of volume lead to different degenerate states, one can
use dynamical input as additional information. After all, the
singularity issue is a dynamical problem, and isolating classical
geometries or quantum states corresponding to a singularity has to be
done with knowledge of the evolution equations. In a canonical
quantization, the evolution equations are given by the Hamiltonian
constraint operators. They do indeed contain the volume operator, but
only in combination with holonomies as a commutator \cite{QSDI}. Where
explicit evolution equations have been obtained
\cite{IsoCosmo,Bohr,HomCosmo,SphSymmSing}, their coefficients are
determined by matrix elements of these commutators. It is the
potential vanishing of these coefficients which signals the
possibility of a singularity in the quantum evolution, and knowing
these coefficients is a prerequisite for understanding the removal of
singularities at the quantum level. The conclusion is that, from a
dynamical point of view, one should not look at states annihilated by
the basic volume operator, but at states annihilated by commutators
between the volume operator and holonomies as they appear in the
constraints. These commutators also make use of the relation
(\ref{commut}) and are thus much closer to expressions used for
inverse volume. In particular, all such commutators, whether they
quantize positive or inverse powers of volume or other metric
components, have generically\footnote{There could be accidental
  cancellations in explicit expressions of matrix elements which could
  imply slightly different kernels in some cases.  This would depend
  very sensitively on properties of the volume spectrum and show that,
  if a particular operator would have a bigger kernel, the additional
  states would be very special.} the same kernel, smaller than that of
the basic volume operator, since they differ only in containing
different powers of the volume operator.  Issues of unboundedness on
degenerate states, as in \cite{BoundFull}, then do not arise.

\subsection{Degenerate configurations}
\label{Deg}

The situation is simpler in models studied so far since they have
volume eigenstates which are identical to eigenstates of commutators
of volume and holonomies (see also Sec.~\ref{Models}).  To shed more
light on the relation between the full non-Abelian theory and those
models, we study different degenerate configurations in ${\rm U}(1)^3$
and in particular inverse volume operators. Since, e.g., diagonal
homogeneous models can be viewed as special cases of 6-valent ${\rm
  U}(1)^3$-vertices, one can directly see where potential differences
come from.

In SU(2) one typically has objects such as $\tr\tau_I
h[h^{-1},\hat{V}]$ where holonomies in the commutator when acting on
an edge of spin $j$ lead to contributions with higher spin
$j+\frac{1}{2}$ and lower spin $j-\frac{1}{2}$. In a symmetric setting
such as loop quantum cosmology, whose composite operators are modeled
on SU(2) expressions of the full theory, one can take the trace
explicitly and obtain products of the form
$\sin\frac{c}{2}\hat{V}\cos\frac{c}{2}-
\cos\frac{c}{2}\hat{V}\sin\frac{c}{2}$ where $c$ is a connection
component appearing in holonomies $h(c)=\exp(c\tau)=
\cos\frac{c}{2}+2\tau \sin\frac{c}{2}$ \cite{InvScale,IsoCosmo,Bohr}.
Since, e.g., $\cos\frac{c}{2}= \frac{1}{2}(U(c)+U(c)^{-1})$ with
$U(c)=\exp(ic/2)$, the resulting operators are close to the behavior
in the non-Abelian setting because $U$ as a multiplication operator
increases the label $\mu$ of a state $\langle c|\mu\rangle=\exp(i\mu
c/2)$ while $U^{-1}$ decreases it.  Nevertheless, since one can simply
interpret $U(c)=\exp(ic/2)$ as an Abelian holonomy, one can also view
this expression as obtained in an Abelian setting. However, in a
directly Abelianized model one would rather use an expression such as
$iU[U^{-1},\hat{V}]$, simply replacing the non-Abelian holonomies by
Abelian ones \cite{QFTonCSTII}.  The action is then quite different
because an Abelian holonomy either increases the charge or decreases
it, but does not give two such contributions at the same time as
happens with SU(2).

The more symmetric treatment which does model some non-Abelian
behavior is precisely what leads to an inverse volume vanishing on
zero-volume states in isotropic loop quantum cosmology as can easily
be seen from the resulting eigenvalues. In an isotropic model, the
purely Abelian version \cite{BohrADM} results in eigenvalues of the
form $V_{\mu+1}^r-V_{\mu}^r$ with $V_{\mu}\propto |\mu|^{3/2}$ while
the symmetric form of loop quantum cosmology gives
$V^r_{\mu+1}-V^r_{\mu-1}$ as in (\ref{Iso}). At zero volume, $\mu=0$,
the first result is non-zero unlike the latter one.  Thus, while in
loop quantum cosmology the inverse volume is automatically zero on
zero-volume states, not by hand but by following constructions of the
full theory, the loop inspired treatment of \cite{BohrADM} and also
the Abelianized version of \cite{BoundFull} can give non-zero results.
(Ref.\ \cite{BoundFull} also considers vertices of higher valence
where even the symmetric treatment can give non-zero results on zero
volume, as discussed below. However, those higher valent
configurations are not contained in a homogeneous setting.) This
demonstrates the importance of following full expressions as closely
as possible in models, as advocated in loop quantum cosmology. While
basic operators can directly be derived from the full theory
\cite{SymmRed,SphSymm,LQC}, which is the crucial difference between
loop quantum cosmology and an ordinary minisuperspace quantization,
this is more complicated (and so far unfinished) for composite
operators such that for them guidance from the full theory must be
employed in their constructions.

We can now discuss the issue of zero-volume states in the ${\rm
  U}(1)^3$ setting, still using our 6-vertex but now with unrestricted
charges collected in a $3\times 3$ matrix $n^I_i$. The volume of such
a configuration is given by $V_n\propto \sqrt{\left|\det n\right|}$,
such that zero-volume configurations span an 8-parameter subset of the
9-dimensional configuration space.

Following loop quantum cosmology, powers of metric components
expressed through commutators will take eigenvalues of the form
\[
 {}^{(r)}e^{i_0}_{I_0}\propto V^r_{n^I_i+
  \delta^I_{I_0}\delta_i^{i_0}}- V^r_{n^I_i-
  \delta^I_{I_0}\delta_i^{i_0}}
\]
where only one coefficient $n^{I_0}_{i_0}$ changes in the volume
labels. For any matrix $n^I_i$, not just diagonal ones, this can
easily be seen to be zero using multi-linearity of the determinant
since the volume eigenvalues depend only on the absolute value of the
determinant of $n^I_i$. Thus, a whole row of components
${}^{(r)}e^{i_0}_{I_0}$ vanishes in those cases and so do all volume
or inverse volume operators $\det{}^{(r)}e^i_I$ constructed from them.
The main difference to diagonal models is that cancellations as in
(\ref{Vdiag}) do not happen; but this is not problematic since they
have not been made use of in loop quantum cosmology, anyway.

The advantage of using a 6-vertex of the above form is that the
geometrical meaning of its labels is clear from the identification of
$n^I_i$ with eigenvalues of densitized triad components $p^I_i$
through flux operators. As we have demonstrated, boundedness can be a
direct consequence of effectively Abelian behavior, but it is not
directly related to symmetry in the form of diagonalization or
isotropy which only leads to more special matrices $n^I_i$.  The
situation for a 6-vertex in SU(2) of a similar type, which could still
be interpreted as corresponding to a homogeneous model, is
unfortunately more complicated and so far unknown.\footnote{The
  unboundedness on non-Abelian 3-vertices suggests that also regular
  6-vertices with equal spins on opposite edges will have unbounded
  expectation values of inverse volume.}

One can generalize the Abelian setting by introducing more edges (and
thus more degrees of freedom, albeit still finitely many ones).
Volume eigenvalues are then given not by a single determinant but by a
sum of determinants of different matrices (for each non-planar triple
of edges). In many cases one can reduce this to a single determinant
using multi-linearity and gauge invariance, but in general this is not
possible and there are explicit examples with unboundedness \cite{JB}.
However, it is not clear what this means since also the geometrical
interpretation of labels would be lost: We would have more independent
labels than the nine values contained locally in a classical triad.
Moreover, such an Abelian situation cannot come from a geometrically
motivated symmetry reduction where any vertex which is more than
6-valent (or 6- or less-valent but without straight edges) could not
be embedded in a ${\rm U}(1)^3$-vertex.\footnote{Effectively Abelian
  behavior occurs when all independent holonomies have orthogonal
  internal directions (see, e.g., \cite{SphSymm}) such as
  $e^{c_{(I)}\tau_I}$ in diagonal homogeneous models. This is no
  longer possible if there are new independent connection components
  from additional edges.}

A different generalization brings us to inhomogeneities: So far we
assumed edge labels to be identical for opposite edges of the
6-vertex, which can be interpreted as diagonal homogeneous models.
When this condition is dropped, one leaves homogeneity but the
configurations can still be interpreted as vertices of inhomogeneous
models. (Indeed, also in the SU(2)-setting the 3-vertices of
\cite{BoundFull} can be viewed as extreme forms of such inhomogeneous
6-vertices with three vanishing labels.) For instance, if we have an
Abelian vertex with only one such inhomogeneous edge with opposite
labels $k_{\pm}$ and of diagonal form, it can be viewed as a general
vertex of a polarized cylindrical wave model \cite{SphSymm}.  Volume
eigenvalues are then of the form $V_{k_{\pm},\nu,\mu}\propto
\sqrt{\left|k_++k_-\right|\mu\nu}$.  The behavior of inverse volume
can easily be seen to be unchanged since we simply replace a label by
a sum of two labels. Also in non-diagonal cases (which include the
configurations used in \cite{QFTonCSTII} without any symmetry
assumptions), the previous conclusions about boundedness do not
change. Thus, even inhomogeneity does not lead directly to unbounded
behavior.

\subsection{Comparison}
\label{Comp}

From the preceding expressions and discussion it is clear that a
direct technical reason for vanishing inverse volume operators on zero
volume eigenstates lies in volume eigenvalues which depend on the edge
labels only through the absolute value of a multi-linear function.
Examples are the multi-linear functions $\mu$ in the isotropic case or
$\det n$ for a ${\rm U}(1)^3$ 6-vertex. This is, however, not a
necessary condition as demonstrated by the isotropic model where
alternative volume eigenvalues $((|\mu|-1)|\mu|(|\mu|+1))^{3/2}$,
somewhat closer to SU(2) expressions, have been used
\cite{cosmoII,IsoCosmo} which also give zero inverse volume
eigenvalues. Nevertheless, the appearance of multi-linear functions is
quite common, as they can occur e.g.\ in isotropic, diagonal
homogeneous and some inhomogeneous models. It is, however, difficult
to find a geometrical reason for unboundedness since many different
possibilities are realized. Linking unbounded behavior to, say,
inhomogeneities is impossible because there are (non-diagonal)
homogeneous models whose volume eigenvalues do not depend on edge
labels through a multi-linear function \cite{cosmoII}, but also
inhomogeneous models which do have such a dependence
\cite{SphSymmHam}.

There are different ways to break the multi-linearity by considering
configurations with additional parameters. The simplest way is by
adding more edges and their labels as discussed in the Abelian
setting. Sometimes there are also alternative volume operators in
models, e.g.\ \cite{SphSymmVol}, which lead to a different dependence
on labels and thus can possibly change the qualitative behavior of
eigenvalues. This situation is similar to the full theory in that the
volume operator does not commute with its commutator with
holonomies. Finally, there are models or the full theory where
non-Abelian effects play a role.  Here, new parameters are in fact
provided by using general non-Abelian holonomies while in symmetric
models often only the Killing norm of su(2) elements is relevant.
Additional correction terms are automatically switched on when
non-Abelian degrees of freedom are added, as we already saw in
Sec.~\ref{Abel} with additional connection dependent terms in
identities such as (\ref{commut}).

Additional parameters appear only if one takes into account more
degrees of freedom than realized in a given symmetric model.  However,
the physical role of these additional degrees of freedom is not always
clear since the geometrically and physically relevant ones have
already been picked out by the symmetry reduction. With additional
degrees of freedom, one thus has always a new physical situation which
cannot immediately be compared to the original, e.g.\ symmetric one.
It is important to keep in mind that, in contrast to a minisuperspace
quantization, the selection of degrees of freedom in loop quantum
cosmology is not done solely by hand.  Common to a minisuperspace
reduction is the choice of a symmetry to be imposed, which specifies
the physical situation of interest and the corresponding reduced
classical phase space. Symmetric states, in the connection
representation, are then defined as distributions in the full theory
whose support contains all relevant invariant connections as a dense
subset \cite{SymmRed}. Since the support of a distribution is by
definition a closed set, it is not automatic that symmetric states are
not supported on some connections which do not appear as invariant
ones at the classical level. At this point, crucial information from
the full configuration space of generalized connections enters the
basic construction of symmetric models. As the analysis reveals, the
classical set of invariant connections is indeed extended to a bigger
(compactified) space of generalized invariant connections, which can
also be introduced at the minisuperspace level \cite{Bohr}. However,
one can prove \cite{SymmRed} that this space of generalized invariant
connections is a closed subset of the full space of generalized
connections and indeed the support of symmetric distributions.  Thus,
reducing at the kinematical quantum level does not add new degrees of
freedom, which shows that kinematically the symmetry reduction
followed in loop quantum cosmology is consistent within the full
theory.  This justifies the treatment along the lines of a
minisuperspace quantization, an issue which could never have been
addressed without a detailed relation between models and the full
theory. More complicated, and still open, is the question of how the
dynamics of models and the full theory are related.

\section{Evolution}
\label{Evol}

In addition to geometrical, static properties of degenerate
configurations one also needs to know their dynamical role as recalled
in Sec.~\ref{General}.  This is achieved in models by using difference
equations in internal time representing the Hamiltonian constraint in
the triad representation, which show how degenerate configurations are
approached during physical evolution.  The existence of such
difference equations which are globally defined, i.e.\ on the full
mini- or midisuperspace of the model, relies on the removal of some
non-Abelian effects.  General non-Abelian behavior makes fluxes,
representing triad components, non-commutative\footnote{This is to be
distinguished from the behavior discussed in Sec.~\ref{Abel} as a
consequence of (\ref{commut}). In this case, correction terms lead to
non-commuting operators even when they correspond to metric components
smeared along the same surface.} such that not all of them can be
sharp at the same time and no triad representation exists
\cite{NonCommFlux}.  Nevertheless, local versions of the difference
equation, as indeed used in inhomogeneous models \cite{SphSymmSing},
can still be possible.  In models, as discussed before, one can then
see that unboundedness, even if it occurs such as in anisotropic
models, is no obstruction to non-singular evolution. Originally, a
non-symmetric ordering had been used which (together with its
associated dynamical initial conditions
\cite{DynIn,Essay} at least in their most straightforward incarnation)
would not work with non-zero matter densities at zero volume. Such an
ordering, however, is already ruled out in inhomogeneous models
\cite{SphSymmSing} even in the vacuum case. With the symmetric
ordering currently preferred it is not necessary for non-singular
behavior that inverse volume expressions are zero or bounded on
zero-volume states \cite{IsoCosmo,BHInt}.

\subsection{Full theory}

We can now ask what dynamical effects one can expect from properties
of inverse volume in the full theory.  So far the observed
unboundedness of inverse volume on tri-valent vertices of the full
theory has no direct implications for the singularity removal
mechanism of loop quantum cosmology. Those states can be considered
zero-volume states (even though also this depends on quantization
ambiguities, see Sec.~\ref{Abel}) but it is not clear how they occur
in a physical transition. For this, one would also have to consider
states of non-zero volume which in some sense are close to tri-valent
vertices. This could be conceivable to do in a 4-valent setting, or on
a 6-vertex using \cite{Loll:Simply}, where one would have to determine
the volume spectrum on the intertwiner space.  Parity reversal, which
is important in models, will then be controlled not by spins of outer
edges but by intertwiner labels.  Changing from positive to negative
eigenvalues of $\det E$ would not affect outer edge spins and thus not
result in an intermediate 3-vertex. This could only arise after using
the full Hamiltonian constraint, which does change edge spins.  After
finding suitable such transitions, one has to see if they are required
by the dynamical evolution and whether or not evolution breaks down.
This could possibly be done locally by a difference equation with an
intertwiner spin as clock variable even though due to the non-Abelian
nature no global difference equation for the constraint on superspace
exists.

A mechanism alternative to using dynamical information from difference
equations, which would be much more complicated to obtain in the full
theory, was suggested by using coherent states and following their
behavior when they are peaked on geometries of ever smaller volume
\cite{BoundCoh}. One can then follow the expectation value of matter
energy, which indeed in a ${\rm U}(1)^3$ calculation has been shown to
remain bounded provided that fluxes are bounded along the classical
evolution (as realized, e.g., for vacuum Bianchi class A models even
when approaching the classical singularity). This coincides with the
expectation from anisotropic models.\footnote{Indeed, a similar
procedure developed in models consists in using effective equations
\cite{Inflation,Time,DiscCorr,SemiClassEmerge,Perturb,Josh}. This has
been used, e.g., to study modifications to classical chaos in the
Bianchi IX model \cite{Mixmaster} where the situation is comparable to
that in \cite{BoundFull,BoundCoh}: the effective evolution is governed
by an {\em unbounded} curvature potential which, however, is {\em
bounded} from above at {\em fixed volume} and in particular {\em along
the evolution} \cite{NonChaos,ChaosLQC}.} But it also shows the
potential dangers of full considerations with our current limited
knowledge: Somewhat hidden in the calculations of \cite{BoundCoh} is
the {\em assumption that fluxes remain bounded} during the approach to
the singularity. If this is not the case, there is no boundedness
result from the ${\rm U}(1)^3$ calculation. Relevant for boundedness
is not the fact that coherent states are used in contrast to volume
eigenstates in the SU(2) calculation, but the fact that {\em
dynamical} information is put in as a key additional assumption. The
danger is that this assumption is based on classical dynamical
properties which moreover are known only in symmetric
situations. Thus, although the calculations of \cite{BoundCoh} are
done in the full setting, crucial for boundedness results are
properties of symmetric solutions. It would be a large step forward in
our understanding of singularities if the boundedness of densitized
triad components close to classical singularities could be shown more
generally. This is also of importance for the mechanism of loop
quantum cosmology, as emphasized in \cite{HomCosmo}.

A similar calculation for SU(2) would be much more involved, but to
some degree one can rely on arguments that expectation values in ${\rm
U}(1)^3$ coherent states are close to what one gets with SU(2).
Nevertheless, potential differences can occur particularly when
unbounded behavior is to be studied. In ${\rm U}(1)^3$, the
expectation value is independent of the connection where the coherent
state is peaked since it appears only in a phase factor, but this is
not the case in a non-Abelian setting. Moreover, as discussed above,
in a non-Abelian calculation there are correction terms to the basic
identity which are small for small connection components. In contrast
to an Abelian calculation we thus expect an expectation value of
inverse volume operators in a coherent state which is also connection
dependent, and moreover correction terms which in general are small
only for small connections entering the coherent state (a coherent
state requires a gauge choice to specify the peak position, thus making
it meaningful to talk about small connections). Accordingly, the
Abelian calculation should be reliable in semiclassical regimes, but
not necessarily for strong curvature. While one can give arguments for
the boundedness\footnote{As in the ${\rm U}(1)^3$ calculation of
\cite{BoundCoh} and as discussed above, this only refers to
boundedness of the expectation value as a function of spin labels {\em
provided that all fluxes remain bounded}. These arguments can thus be
used to illustrate the expected effective semiclassical behavior, but
say nothing about boundedness of operators and the more complicated
issue of extending evolution beyond a classical singularity.} of
expectation values of inverse volume even in SU(2) coherent states
\cite{TT}, based on the fact that holonomies are bounded functions of
the connection, correction terms are important for more detailed
aspects and in particular cosmological phenomenology.  Since
classically connection components, entering non-Abelian correction
terms, become arbitrarily large when a singularity is approached, the
${\rm U}(1)^3$ truncation of the full theory has to be supported by
additional arguments.  Similarly, potential correction terms to
symmetric models and their dynamical role are still to be studied.

\subsection{Models}
\label{Models}

A general extension of wave functions beyond classical singularities
has been found to be realized in all models so far where singularities
have been investigated
\cite{Sing,Bohr,HomCosmo,Spin,Modesto,BHInt,SphSymmSing}. We now give
some details in the isotropic case which also show how bringing in
additional parameters of su(2) elements could potentially change the
picture. Similarly to inverse volume operators, the Hamiltonian
constraint \cite{QSDI} is constructed from holonomies and the volume
operator which in the flat case gives \cite{cosmoIII,IsoCosmo}
\[
 \hat{H}\propto \sum_{IJK}\epsilon^{IJK}\tr(h_Ih_Jh_I^{-1}h_J^{-1}
 h_K[h_K^{-1},\hat{V}])\,.
\]
Holonomies are of the form $h_I=\exp(c\Lambda_I^i\tau_i)$ with the
isotropic connection component $c$ and an SO(3) matrix $\Lambda_I^i$
specifying internal directions which are pure gauge.  Using $h_I=\cos
{\textstyle\frac{c}{2}}+2\Lambda_I \sin {\textstyle\frac{c}{2}}$ with
$\Lambda_I:=\Lambda_I^i\tau_i$, the two contributions are
\begin{equation} \label{hols}
 \epsilon^{IJK}h_Ih_Jh_I^{-1}h_J^{-1}=
 8\Lambda_K\sin^2{\textstyle\frac{c}{2}}\cos^2{\textstyle\frac{c}{2}}+
 4\epsilon^{IJK}(\Lambda_I-\Lambda_J)
 \sin^3{\textstyle\frac{c}{2}}\cos {\textstyle\frac{c}{2}} 
\end{equation}
and
\begin{eqnarray} \label{comm}
 h_K[h_K^{-1},\hat{V}]&=& \hat{V}-\cos
 {\textstyle\frac{c}{2}}\hat{V}\sin {\textstyle\frac{c}{2}}- \sin 
 {\textstyle\frac{c}{2}}\hat{V}\cos {\textstyle\frac{c}{2}}\nonumber\\
 &&- 2\Lambda_K (\sin {\textstyle\frac{c}{2}}\hat{V}\cos
 {\textstyle\frac{c}{2}}-\cos 
 {\textstyle\frac{c}{2}}\hat{V}\sin {\textstyle\frac{c}{2}})\nonumber\\
&&- 2\sin {\textstyle\frac{c}{2}}\cos {\textstyle\frac{c}{2}}
[\Lambda_K,\hat{V}]- 
 4\Lambda_K\sin^2{\textstyle\frac{c}{2}}[\Lambda_K,\hat{V}]\,.
\end{eqnarray}

In the isotropic model, $\Lambda_I^i$ commutes with the volume operator
and its columns are orthogonal to each other. The constraint then
reduces to
\[
 \hat{H}_{\rm iso}\propto \sin^2
 {\textstyle\frac{c}{2}}\cos^2{\textstyle\frac{c}{2}}(\sin
 {\textstyle\frac{c}{2}}\hat{V}\cos {\textstyle\frac{c}{2}}-\cos 
 {\textstyle\frac{c}{2}}\hat{V}\sin {\textstyle\frac{c}{2}})
\]
(or its symmetrization) which has been used for the proof of
non-singular behavior \cite{Sing}. Here, one uses the action
\begin{eqnarray}
 \cos {\textstyle\frac{c}{2}}|\mu\rangle &=&
 \frac{1}{2}(|\mu+1\rangle+|\mu-1\rangle)\\ 
 \sin {\textstyle\frac{c}{2}}|\mu\rangle &=&
 -\frac{i}{2}(|\mu+1\rangle-|\mu-1\rangle)  \label{sin}
\end{eqnarray}
and derives a difference equation in the triad representation for a
wave function $|\psi\rangle=\sum_{\mu}\psi_{\mu}|\mu\rangle$.

We can now bring in new parameters by allowing arbitrary $\Lambda_I^i$
except that their norm must still be one (since otherwise $c$ would be
redundant). This allows six rather than three angles in specifying the
internal directions, and three of them remain after removing gauge
freedom.  Analyzing the canonical structure of the new model and the
meaning of the additional parameters (which can be related to shape
parameters following \cite{AshSam}) is beyond the scope of this paper,
but we can already see potential extra terms in the constraint. The
following statements can be interpreted in the context where these
additional parameters are switched on perturbatively such that the
basic representation of an isotropic model remains unchanged except
that states now are also functions of the new angles.

There are two conditions of the isotropic model which are no longer
satisfied: (i) $\tr\Lambda_I\Lambda_K=0$ for $I\not=K$ and (ii)
$[\Lambda_K,\hat{V}]=0$. Dropping the first condition implies that
also the second term in (\ref{hols}) contributes to the constraint
which changes the difference equation. However, there is no crucial
change in structure since the equation remains of the same order (the
number of trigonometric functions in each term is the same).
Leading coefficients of the difference equation could a priori be
vanishing in different places with the new term, but this does not
happen since the new contribution is imaginary (the additional sine
implies an additional factor of $i$ in a difference operator from
(\ref{sin})).  The only change is thus in coefficients becoming
complex.

Dropping the second condition is more important since it leads to a
higher order of difference equations. (Note that we always have
$0=[\Lambda_K^2,\hat{V}]=\Lambda_K\cdot[\Lambda_K,\hat{V}]+
[\Lambda_K,\hat{V}]\cdot\Lambda_K$ using the fact that the $\Lambda_K$
are normalized. But an individual commutator $[\Lambda_K,\hat{V}]$ can
still be non-zero.)  Moreover, the last two terms in (\ref{comm}),
unlike the first two lines, do not commute with the volume operator.
This is different from other homogeneous models where inverse volume
operators commute with the volume operator, but it would be analogous
to behavior in the full theory as noticed in Sec.~\ref{Abel}. Indeed,
the extra terms in (\ref{comm}) can also contribute to non-zero
expectation values of inverse volume operators in zero-volume
eigenstates. If we use
$\hat{O}:=\tr(\Lambda_Kh_K[h_K^{-1},\hat{V}^r])$ for an inverse power,
we will have $\langle0|\hat{O}|0\rangle=0$ in the isotropic zero
volume eigenstate $|0\rangle$ since $\tr[\Lambda_K,\hat{V}]=0$ and
$\langle0|\sin {\textstyle\frac{c}{2}}\cos
{\textstyle\frac{c}{2}}|0\rangle=0$. But for $I\not=K$ we can have
$\langle0|\tr(\Lambda_Ih_K[h_K^{-1},\hat{V}^r])|0\rangle\not=0$ from
the last term in (\ref{comm}) which also corresponds to some inverse
power of volume. The behavior is thus closer to that in the full
theory and provides an explicit example for the role of non-Abelian
effects in unboundedness as discussed in general in Sec.~\ref{Abel}.
Still, features of the isotropic model are recognizable. In
particular, there is a difference equation which, compared to the
isotropic one, is of higher order and has different coefficients.
Deriving implications for the singularity issue requires an
interpretation of new degrees of freedom in $\Lambda$ in relation to
classical metric components as well as an understanding of the
classical singularity structure. This requires a more detailed
analysis of the canonical and geometrical structure, but the above
considerations already show that kinematical properties such as
those discussed in \cite{BoundFull} can be studied in models without
jumping directly to the full theory.

\section{Summary}

Discussing the singularity problem is made complicated by the fact
that there are several different issues to be kept in mind. We have
presented several technical points in this paper, so far mainly in
separation from each other. Before concluding we now collect these
observations.

\subsection{Kinematical vs.\ Dynamical}

Information used for the singularity issue includes
\begin{itemize}
 \item Kinematical properties: eigenvalues of inverse volume or
 intrinsic curvature operators, or their expectation values in zero
 volume states or kinematical coherent states,
\item Dynamical properties: the form of difference equations derived
 from the constraint, effective equations or classical dynamical
 properties in coherent states.
\end{itemize}

As inverse volume may or may not be bounded, its relevance can be seen
only in conjunction with dynamics, e.g.\ from how it appears in an
evolution equation. Kinematically, it does not matter which kinds of
states are used to compute expectation values of inverse volume, and
also in coherent states does one generally still have
unboundedness. The advantage of using coherent states, however, is
that it is easier to bring in dynamical information through the
classical peak position. One can then show that for bounded fluxes
also inverse volume is bounded, requiring an additional assumption
motivated by classical properties of symmetric solutions. So also here
symmetric properties are being used, even though the intention is to
go beyond symmetry. (Strictly speaking, the argument in
\cite{BoundCoh} does not only use available classical information
because it assumes boundedness of {\em all} fluxes around a given
vertex. This still has to be related to classical behavior, in
particular considering the fact that more independent labels are
required for an ${\rm U}(1)^3$ vertex with unbounded inverse volume
than one has locally in a classical triad.) This illustrates the
current state of understanding of singularities not just from the
direction of quantum gravity but even classically. Precise statements do
not exist without assuming at least some degree of symmetry. One can
try to find perturbative behavior around a symmetric solution, but in
particular for the singularity issue this can be very misleading since
it is not clear how perturbations would grow. (Also from a technical
perspective, this has been explicitly observed in
\cite{AnisoPert}.) Difference equations provide alternative dynamical
input where the situation can be decided fully at the quantum
level. Classical information is then required only to identify
classical singularities. So far, however, this procedure is restricted
to symmetric situations.

\subsection{Isotropy, anisotropy, inhomogeneity}

As already mentioned in the beginning, isotropy is very special with
minisuperspace having the same dimension as trajectories. There is
then only one way to approach a classical singularity, and inverse
volume expressions are completely bounded. Anisotropic models are more
general and display almost all of the explicitly known characteristic
behavior of classical singularities. Here, curvature is in general
unbounded on minisuperspace. There are some inhomogeneous models which
are not necessarily more complicated as far as unboundedness is
concerned, but dynamical properties are more difficult to
disentangle. In other models or in the full theory, even the
boundedness issue can be problematic as a consequence of non-Abelian
behavior.

The latter may be a potential source for differences between models
and the full theory, but its relevance requires a better understanding
of non-Abelian degrees of freedom. This is of importance for the role
of additional ambiguities and for finding the relation between
degenerate and singular states. From this point we come to the final
issue:

\subsection{Non-Abelian behavior vs.\ inhomogeneity}

Non-Abelian properties clearly have to be taken into account in the
full theory, but also in several models such as non-diagonal
homogeneous models \cite{cosmoI,cosmoII} or unpolarized cylindrical
waves \cite{CylWaveVol}. Inhomogeneities certainly appear in the
full theory but also in models such as spherical symmetry or
cylindrical waves. Here, one should remember that the role of all
the degrees of freedom provided by loop quantum gravity or
inhomogeneous models is not yet clear; in particular, the relation
between mathematical degrees of freedom and physical inhomogeneities,
i.e.\ fields on a given homogeneous background, is not fully
understood. This is relevant especially when vertices of arbitrarily
high valence are considered in the full theory which provides a
potentially infinite number of degrees of freedom but of undetermined
geometrical role.

This issue is the most difficult to analyze because in general there
is no clear separation between non-Abelian effects, inhomogeneities
and the valence of vertices. High-valent vertices can occur only in
the full setting or in models whose symmetry orbits have at most
dimension one. However, there are Abelian as well as non-Abelian
homogeneous models, and also both kinds of inhomogeneous models.

To avoid having to deal with arbitrarily high valent vertices we can
restrict ourselves to three-valent ones, or six-valent if there are
three straight edges entering and leaving a vertex, since this is
anyway the only calculation available from the full point of view so
far. It also represents the combinatorial situation of models where
singularities have been studied so far. Then, a distinction between
non-Abelian effects and inhomogeneities becomes possible as discussed
in Sec.~\ref{Abel}: While the non-Abelian volume operator does not
commute with inverse volume, the latter then having unbounded
expectation values in states annihilated by the former, volume and
inverse volume do commute in the Abelian case and inverse volume
vanishes on zero volume states. Here, inhomogeneity does not play any
role whatsoever and thus cannot be responsible for unboundedness. In
an Abelian calculation, one obtains unboundedness only at vertices of
high valence for which the geometrical role of their labels is
unknown.

We emphasize again that inverse volume is considered here only because
it is the best understood operator in this context. Otherwise,
unboundedness properties of curvature already arise when introducing
anisotropy, not even inhomogeneities.

\section{Conclusions}

We have demonstrated that, from currently available information, there
is no contradiction between models and the full theory of loop quantum
gravity.  Instead, a consistent picture emerges when all cases,
isotropic as well as anisotropic or inhomogeneous models and the full
theory, are considered. There is a technical difference between
inverse volume in the full theory and similar operators in models, in
that the expectation values in the full theory are unbounded on states
of vanishing volume eigenvalue, while they are zero on such states in
models considered so far. However, in the full theory the distinction
of singular states, in particular their identification with certain
degenerate states, is blurred due to non-Abelian effects. There is
then no crucial difference between a model where curvature is
unbounded {\em close to singular states} (i.e.\ for quantum labels
close to those of a singular state) and the full theory where inverse
volume is unbounded {\em on degenerate states} which can be argued to
be close to, but not necessarily identical with singular states. Any
claim of a contradiction between models and the full theory, based on
these properties, is thus unsubstantiated.

Nevertheless, there are certainly differences by design since models
capture behavior only in a particular, geometrically selected sector
of physical interest. What is clear from the considerations is that so
far inhomogeneities cannot be made responsible for any explicit
discrepancy between models and the full theory. To judge what
implications particular properties of inverse volume operators have,
their geometrical and dynamical roles must be clear before rushing to
conclusions. The latter can be derived, e.g., through difference
equations representing the Hamiltonian constraint or observables. But
also from the classical side one needs to provide knowledge on the
general singularity structure, which becomes exceedingly complicated
when symmetry assumptions are dropped (see, e.g., \cite{NumSing}).

The main issue to be checked in further investigations is implications
of non-Abelian effects not studied so far in the different sectors of
loop quantum gravity (including not only cosmological situations but
also black hole horizons and other models or semiclassical states, all
of which often exploit possible eliminations of non-Abelian terms in
explicit calculations). In models, understanding the meaning of
degenerate configurations is achieved by the selection of an
appropriate sector of the theory displaying the configurations of
interest explicitly. We emphasize again that at this point crucial
information from the full theory enters through distributional states.
The symmetry is specified for the physical context, and then the
relevant quantum degrees of freedom result through derivation.
(Similarly, imposing black hole horizons selects the appropriate
degrees of freedom, again in an effectively Abelian manner, through
physical conditions for an isolated horizon \cite{IHPhase} relating
horizon degrees of freedom to full flux operators.) This suggests that
non-Abelian degrees of freedom are not always crucial physically,
which can also be seen from the fact that diagonal homogeneous models,
which in contrast to non-diagonal ones can effectively be Abelianized,
show the complete behavior of cosmological evolution \cite{AshSam}.

Still, additional correction terms in a non-Abelian situation can
provide characteristic effects.  Better understanding non-Abelian
behavior is thus not just important for the relation between models
and the full theory, i.e.\ as a test of approximations, but can also
provide new physical insights.  The non-commutative behavior of
quantum geometry \cite{NonCommFlux} so far has not been made use of in
cosmological investigations in loop quantum gravity, while
non-commutative geometry itself has given rise to several cosmological
applications (see, e.g., \cite{NonCommInfl,NonCommCMB}).  Combining
loop cosmological phenomenology with non-commutative behavior thus has
the potential of providing further scenarios for the very early
universe.

\section*{Acknowledgements}

The author is grateful to Abhay Ashtekar, Johannes Brunnemann,
Ghanashyam Date and Thomas Thiemann for discussions.


\end{document}